\documentclass[prd,aps,showpacs,twocolumn,
               superscriptaddress,amsfonts,eucal]{revtex4}
\usepackage{amsmath}
\usepackage{latexsym}
\usepackage{graphicx}
\usepackage{epstopdf}
\begin{document}

%%%%%%%%%%%%%%%%%
%%%   TITLE   %%%
%%%%%%%%%%%%%%%%%

\title{Scalar functions for wave extraction in numerical relativity} 

%%%%%%%%%%%%%%%%%%%
%%%   AUTHORS   %%%
%%%%%%%%%%%%%%%%%%%

\author{Andrea~Nerozzi}
\affiliation{
Center for Relativity, University of Texas at
Austin, Austin TX 78712-1081, USA
}

%%%%%%%%%%%%%%%%
%%%   DATE   %%%
%%%%%%%%%%%%%%%%

\date{\today}

%%%%%%%%%%%%%%%%%%%%
%%%   ABSTRACT   %%%
%%%%%%%%%%%%%%%%%%%%

\begin{abstract}
Wave extraction plays a fundamental role in the 
binary black hole simulations currently performed in numerical
relativity. Having a well defined procedure for wave extraction, which matches simplicity with efficiency,
is critical especially when comparing waveforms from different simulations. 
Recently, progress has been made in defining a general technique which 
uses Weyl scalars to extract the gravitational wave signal, through
the introduction of the {\it quasi-Kinnersley tetrad}. This procedure
has been used successfully in current numerical simulations; however, 
it involves complicated calculations. The work in this
paper simplifies the procedure
by showing that the choice of the {\it quasi-Kinnersley tetrad} is reduced
to the choice of the time-like vector used to create it. The space-like
vectors needed to complete the tetrad are then easily identified,
and it is possible to write the
expression for the Weyl scalars in the right tetrad, 
as simple functions of the
electric and magnetic parts of the Weyl tensor. 
\end{abstract}

%%%%%%%%%%%%%%%%
%%%   PACS   %%%
%%%%%%%%%%%%%%%%


%%%%%%%%%%%%%%%%%%%%%
%%%   MAKETITLE   %%%
%%%%%%%%%%%%%%%%%%%%%

\maketitle

%%%%%%%%%%%%%%%%%%%%%%%%
%%%   INTRODUCTION   %%%
%%%%%%%%%%%%%%%%%%%%%%%%

\section{Introduction}
\label{sec:introduction}
Weyl scalars are promising tools for wave extraction in numerical relativity, and 
are being used for this purpose in current binary black hole simulations.
As has been shown in 
\cite{beetle-2005-72,nerozzi-2005-72,nerozzi-2006-73,nerozzi-2006-,burko-2007-},
the critical issue is choosing the right null tetrad in which to calculate the Weyl 
scalars. 
The notion of a {\it quasi-Kinnersley frame} was introduced, which leads to a 
robust technique for their calculation. However, in spite of the relevant result, the procedure
is complicated in practice, although it has been implemented 
with success in \cite{campanelli-2006-73b,campanelli-2006-,
mCcLyZ06a,mCcLpMyZ06, campanelli-2006-74, campanelli-2006-74b,baker-2007-}.
This paper 
gives a fully well-defined and simplified procedure, whose results are completely 
analytical and require no further numerical calculations. The pivotal step
is the identification of a ``preferred'' time-like observer. Then, one can find simple
and well-defined expressions for the scalars
in the {\it quasi-Kinnersley frame}. 

%%%%%%%%%%%%%%%%%%%%%%%%%%%%%%
%%%    WEYL SCALARS   %%%
%%%%%%%%%%%%%%%%%%%%%%%%%%%%%%

\section{Weyl scalars}
\label{sec:weyl}

Weyl scalars are defined as

\begin{subequations}
\label{eqn:weylscalars}
\begin{eqnarray}
\Psi_0 &=& C_{abcd}\ell^am^b\ell^cm^d, \label{eqn:weylscalars0} \\
\Psi_1 &=&C_{abcd}\ell^an^b\ell^cm^d, \label{eqn:weylscalars1} \\
\Psi_2 &=&C_{abcd}\ell^am^b\bar{m}^cn^d, \label{eqn:weylscalars2} \\
\Psi_3 &=& C_{abcd}\ell^an^b\bar{m}^cn^d, \label{eqn:weylscalars3} \\
\Psi_4 &=& C_{abcd}n^a\bar{m}^bn^c\bar{m}^d, \label{eqn:weylscalars4} 
\end{eqnarray}
\end{subequations}
where $C_{abcd}$ is the Weyl tensor, and 
$\ell^a$, $n^a$, $m^a$ and $\bar{m}^a$ comprise a null tetrad with two
real and two complex conjugate null vectors,
such that $\ell^an_a=-1$ and $m^a\bar{m}_a=1$. 

The correct Weyl scalar calculation for wave extraction 
is the one performed in the null tetrad which converges to the Kinnersley tetrad \cite{Kinnersley69,Teukolsky73} when the space-time 
approaches a Petrov type D space-time (a particular algebraic class of space-times to which
Schwarzschild and Kerr belong). This tetrad has been dubbed a {\it quasi-Kinnersley
tetrad}.
Refs \cite{beetle-2005-72,nerozzi-2005-72,nerozzi-2006-73,nerozzi-2006-,burko-2007-} show that the {\it quasi-Kinnersley tetrad} belongs
to a group of tetrads called the {\it quasi Kinnersley frame}, whose tetrads are connected
to each other by spin/boost (type III) transformations. One possible
{\it quasi-Kinnersley frame} was found to be
one of the three transverse frames where $\Psi_1=\Psi_3=0$.

The existence of transverse frames has already been studied in \cite{nerozzi-2005-72} and
in particular it has been shown that for an algebraically general space-time (Petrov type I)
three transverse frames {\it always} exist, while for algebraically special space-times
(Petrov types II,D,III,N) the number of transverse frames varies depending on the algebraic type. Hereafter
we will however assume to be dealing with Petrov type I space-times, which are the ones
normally considered in a numerical simulation.

In this article, Weyl scalars are expressed within the context of the standard 3+1
decomposition of Einstein's equations. A space-like foliation
of the space-time is introduced, with $N^a$ the time-like unit normal to the 
space-like hypersurfaces.
On each hypersurface, we introduce a set of three orthonormal vectors, 
referred to as $u^a$, $x^a$ and $y^a$. 
The set of four vectors -- one time-like, the other three space-like --  leads to a straightforward
definition of the null tetrad:

\begin{subequations}
\label{eqn:nullvectors}
\begin{eqnarray}
\ell^a &=& 2^{-\frac{1}{2}}\left(N^a-u^a\right), \label{eqn:nullvectors1} \\
n^a &=&  2^{-\frac{1}{2}}\left(N^a+u^a\right), \label{eqn:nullvectors2} \\
m^a &=&  2^{-\frac{1}{2}}\left(x^a-iy^a\right), \label{eqn:nullvectors3} \\
\bar{m}^a &=&  2^{-\frac{1}{2}}\left(x^a+iy^a\right). \label{eqn:nullvectors4}
\end{eqnarray}
\end{subequations}

%The three space-like vectors constitute and right handed orthonormal tria, i.e.
%they satisfy the additional condition

%\begin{equation}
%\label{eqn:righthanded}
%\epsilon_{abc}u^ax^by^c = 1,
%\end{equation}
%
%where $\epsilon_{abc}$ is the three-dimensional Levi-Civita tensor.

The Cauchy foliation of the space-time allows us to define
the electric and magnetic parts of the Weyl tensor as

\begin{subequations}
\label{eqn:electricandmagnetic}
\begin{eqnarray}
E_{ac}&=&-C_{abcd}N^bN^d, \label{eqn:electric} \\
B_{ac}&=&-\frac{1}{2}{\epsilon_{ab}}^{mn}C_{mncd}N^bN^d \label{eqn:magnetic},
\end{eqnarray}
\end{subequations}
where ${\epsilon_{ab}}^{mn}$ is the four dimensional Levi-Civita tensor.
The final expression for the Weyl scalars is then given by

\begin{subequations}
\label{eqn:weylele}
\begin{eqnarray}
\Psi_0 &=& -\left(E_{bc}-iB_{bc}\right)m^bm^c, \label{eqn:weylscalars0e} \\
\Psi_1 &=&2^{-\frac{1}{2}}\left(E_{bc}-iB_{bc}\right)m^bu^c, \label{eqn:weylscalars1e} \\
\Psi_2 &=&-2^{-1}\left(E_{bc}-iB_{bc}\right)u^bu^c, \label{eqn:weylscalars2e} \\
\Psi_3 &=& -2^{-\frac{1}{2}}\left(E_{bc}-iB_{bc}\right)\bar{m}^bu^c, \label{eqn:weylscalars3e} \\
\Psi_4 &=& -\left(E_{bc}-iB_{bc}\right)\bar{m}^b\bar{m}^c. \label{eqn:weylscalars4e} 
\end{eqnarray}
\end{subequations}

Eq.~(\ref{eqn:weylele}) is our starting point for determining the correct
null tetrad for computing the Weyl scalars. Specifically, our
{\it quasi-Kinnersley tetrad}
procedure will identify the optimal $u^a$, $x^a$ and $y^a$ for wave extraction,
assuming the right $N^a$ time-like vector has been chosen. 

%%%%%%%%%%%%%%%%%%%%%%%%%%%%%%
%%%    THE TRANSVERSE FRAME %%%
%%%%%%%%%%%%%%%%%%%%%%%%%%%%%%

\section{The transverse frames}
\label{sec:trans}

First, we calculate the frame where $\Psi_1=\Psi_3=0$.
Using  
Eq.~(\ref{eqn:weylscalars1e}) and (\ref{eqn:weylscalars3e}) it can be easily shown that this condition corresponds
to 

\begin{subequations}
\label{eqn:transcond}
\begin{eqnarray}
E_{bc}u^bx^c = 0, \label{eqn:transcond1} &
E_{bc}u^by^c = 0, \label{eqn:transcond2} \\
B_{bc}u^bx^c = 0, \label{eqn:transcond3} &
B_{bc}u^by^c = 0. \label{eqn:transcond4}
\end{eqnarray}
\end{subequations}
The four equations written in Eq.~(\ref{eqn:transcond}) can be reduced to two
by noticing that a spin transformation rotates the vectors $x^a$ and $y^a$ in
their plane, leaving the condition $\Psi_1=\Psi_3=0$ unaltered. Thus, an equivalent
way of writing  Eq.~(\ref{eqn:transcond}) is

\begin{eqnarray}
\label{eqn:transcond2}
E_{bc}u^bw^c = 0, &
B_{bc}u^bw^c = 0,
\end{eqnarray}
where $w^a$ is a generic vector lying in the plane perpendicular to $u^a$.
The only way Eq.~(\ref{eqn:transcond2}) can hold for any vector $w^a$ perpendicular
to $u^a$ is by having both products $E_{bc}u^b$ and $B_{bc}u^b$ parallel to
$u^a$, such that

\begin{eqnarray}
\label{eqn:transcond3}
E_{bc}u^b=E_{\bf u} u_c, &
B_{bc}u^b=B_{\bf u} u_c,
\end{eqnarray}
This means that  the vector $u^a$ {\it is an eigenvector common to the electric and magnetic
parts of the Weyl tensor}, $E_{\bf u}$ and $B_{\bf u}$ being the
corresponding eigenvalues. The three orthonormal common eigenvectors 
correspond to the three possible transverse frames. 

It is evident that the
electric and magnetic components of the Weyl tensor cannot have their eigenvectors
in common for any choice of the time-like vector $N^a$ used to calculate
them. On the other hand, we know that for a specific choice of $N^a$, this
condition must hold, as we know that transverse frames where
$\Psi_1=\Psi_3=0$ do indeed exist. 
This means that there is a ``preferred'' time-like
observer who sees those eigenvectors aligned.
We do not know yet how to identify unequivocally such an observer, but it is 
clear that this is the observer to choose in order to have
meaningful physical results. 

Assuming to have chosen the right time-like
vector for the calculation of $E_{ab}$ and $B_{ab}$, and
denoting 
$\left\{E_{\bf u}, E_{\bf x}, E_{\bf y}\right\}$ the eigenvalues of the electric part, 
and $\left\{B_{\bf u}, B_{\bf x}, B_{\bf y}\right\}$ the eigenvalues of the 
magnetic part, we easily obtain (using Eq.~(\ref{eqn:weylscalars4e})) the
expression for $\Psi_4$:

\begin{equation}
\label{eqn:psi4trans}
\left(\Psi_4\right)_{TF} = \frac{E_{\bf y} - E_{\bf x}}{2}+ i\frac{B_{\bf x}-B_{\bf y}}{2}.
\end{equation}
Eq.~(\ref{eqn:psi4trans}) indicates that to compute $\Psi_4$ in the transverse frame, one only needs
to know the eigenvalues of the electric and magnetic parts of the Weyl tensor.

\section{Eigenvalues of the electric and magnetic parts}
\label{sec:eigen}

In this section, we calculate the expressions for the eigenvalues of the electric part of
the Weyl tensor. The procedure
for the magnetic part is identical. The equation for the eigenvalues is given by

\begin{equation}
\label{eqn:eigen}
\lambda^3-\frac{E_I}{2} \lambda-\frac{E_J}{3} = 0,
\end{equation}
where $E_I$ and $E_J$ are defined as

\begin{eqnarray}
\label{eqn:invariants}
E_I = E_{ab}E^{ab}, &
E_J = {E_a}^b{E_b}^c{E_c}^a. 
\end{eqnarray}

The three eigenvalues can be written as

\begin{subequations}
\label{eqn:eigenvalues}
\begin{eqnarray}
E_{\bf u} &=& -\left(\mathcal{E}+\mathcal{E}^*\right),  \label{eqn:E_u} \\
E_{\bf x} &=& -\left(e^{\frac{2i\pi}{3}}\mathcal{E}+e^{\frac{4i\pi}{3}}\mathcal{E}^*\right),  \label{eqn:E_x} \\
E_{\bf y} &=& -\left(e^{\frac{4i\pi}{3}}\mathcal{E}+e^{\frac{2i\pi}{3}}\mathcal{E}^*\right),  \label{eqn:E_y} 
\end{eqnarray}
\end{subequations}
where $\mathcal{E}$ is a complex number to be determined. This construction guarantees
that the sum $E_{\bf u}+E_{\bf x}+E_{\bf y}$ vanishes, as is expected from the trace-free property of the
electric part of the Weyl tensor. 
It furthermore guarantees that all three eigenvalues are real as expected
from the symmetric property of $E_{ab}$. Using this technique, we only need to calculate
the real and imaginary parts of $\mathcal{E}$, which constitute the two degrees of freedom
of the problem. Substituting Eq.~(\ref{eqn:eigenvalues}) into Eq.~(\ref{eqn:eigen}), we
obtain the following relations:

\begin{subequations}
\label{eqn:eigenres}
\begin{eqnarray}
E_I &=& 6\mathcal{E}\mathcal{E}^*, \label{eqn:eqI} \\
E_J &=& -3\left[\mathcal{E}^3+\left(\mathcal{E}^*\right)^3\right]. \label{eqn:eqJ}
\end{eqnarray}
\end{subequations}

Writing the expression for $\mathcal{E}$ in terms of modulus and phase,
$\mathcal{E} = \left|\mathcal{E}\right|e^{i\Theta_{\mathcal{E}}}$, 
the following expressions are derived from Eq.~(\ref{eqn:eigenres}):

\begin{subequations}
\label{eqn:eigenres2}
\begin{eqnarray}
\left|\mathcal{E}\right| &=& \sqrt{\frac{E_I}{6}} = \sqrt{\frac{E_{ab}E^{ab}}{6}}, \label{eqn:eqmod} \\
\Theta_{\mathcal{E}} &=& \frac{1}{3}\arccos \left[-\sqrt{6}\left(\frac{E_{ab}{E^b}_cE^{ca}}
{\left[E_{ab}E^{ab}\right]^{\frac{3}{2}}}\right)\right]. \label{eqn:eqtheta}
\end{eqnarray}
\end{subequations}
The equations for $\mathcal{B}$ are similar:

\begin{subequations}
\label{eqn:eigenres2B}
\begin{eqnarray}
\left|\mathcal{B}\right| &=& \sqrt{\frac{B_{ab}B^{ab}}{6}}, \label{eqn:eqmodB} \\
\Theta_{\mathcal{B}} &=& \frac{1}{3}\arccos \left[-\sqrt{6}\left(\frac{B_{ab}{B^b}_cB^{ca}}
{\left[B_{ab}B^{ab}\right]^{\frac{3}{2}}}\right)\right]. \label{eqn:eqthetaB}
\end{eqnarray}
\end{subequations}

Eq.~(\ref{eqn:psi4trans}), (\ref{eqn:eigenvalues}), (\ref{eqn:eigenres2}) 
and (\ref{eqn:eigenres2B}) give the expression for the real and imaginary part of
$\Psi_4$ in the transverse frames as

\begin{subequations}
\label{eqn:psi4tf}
\begin{eqnarray}
\left(\Psi_4\right)_{TF}^{\Re} &=& -\sqrt{3}\left|\mathcal{E}\right|\sin\left(\Theta_{\mathcal{E}}+
\frac{2k\pi}{3}\right), \\
\left(\Psi_4\right)_{TF}^{\Im} &=& \sqrt{3}\left|\mathcal{B}\right|\sin\left(\Theta_{\mathcal{B}}+
\frac{2k\pi}{3}\right), 
\end{eqnarray}
\end{subequations}
where $k$ is an integer that can assume the values $\{-1,0,1\}$, corresponding to the
three different transverse frames.
For later convenience, the expression of $\Psi_2$ in 
this same frame is also given:

\begin{subequations}
\label{eqn:psi2tf}
\begin{eqnarray}
\left(\Psi_2\right)_{TF}^{\Re} &=& \left|\mathcal{E}\right|\cos\left(\Theta_{\mathcal{E}}+
\frac{2k\pi}{3}\right), \\
\left(\Psi_2\right)_{TF}^{\Im} &=&-\left|\mathcal{B}\right|\cos\left(\Theta_{\mathcal{B}}+
\frac{2k\pi}{3}\right), 
\end{eqnarray}
\end{subequations}

For this particular tetrad choice $\Psi_0=\Psi_4$.

\section{The single Kerr black hole limit}
\label{sec:qkframe}

Eq.~(\ref{eqn:psi4tf}) is the expression
for $\Psi_4$ in the three transverse frames; 
the validity of this equation is verified below by calculating
the expression for a single Kerr black hole and obtaining
the expected results for the Weyl scalars. Boyer-Lindquist coordinates are used
for the calculation.

The expressions for the electric and magnetic components of the Weyl tensor are 
dependent on the time-like $N^a$ chosen for their calculation.
For now, the correct $N^{\mu}$ is taken to be that given by the Kinnersley tetrad, whose
expression in Boyer-Lindquist coordinates is 

\begin{equation}
\label{eqn:normvector}
N^{a}=\frac{1}{\sqrt{2}}\left[\frac{r^2+a^2}{\Omega},\frac{2\Sigma-\Delta}{2\Sigma},0,\frac{a}{\Omega}\right],
\end{equation}
where $\Omega=\frac{2\Sigma\Delta}{2\Sigma+\Delta}$, 
$\Sigma=r^2+a^2\cos^2\theta$ and $\Delta=r^2+a^2-2Mr$,
$M$ is the mass and $a$ the rotation parameter of the black hole.

The invariant quantities defined in Eq.~(\ref{eqn:invariants}) are then given by

\begin{subequations}
\label{eqn:EBinvaria}
\begin{eqnarray}
E_I &=& \frac{6M^2r^2\left(r^2-3a^2\cos^2\theta\right)^2}{\left(r^2+a^2\cos^2\theta\right)^{6}},
\label{eqn:Einvaria} \\ 
B_I &=& \frac{6M^2a^2\cos^2\theta\left(3r^2-a^2\cos^2\theta\right)^2}{\left(r^2+a^2\cos^2\theta\right)^{6}},
\label{eqn:Binvaria}  
\end{eqnarray}
\end{subequations}
while $E_J = - \frac{E_I^{\frac{3}{2}}}{\sqrt{6}}$ and $B_J- \frac{B_I^{\frac{3}{2}}}{\sqrt{6}}$.
Substituting Eq.~(\ref{eqn:invariants}) into Eq.~(\ref{eqn:eigenres2}) and 
(\ref{eqn:eigenres2B})
gives the following result:

\begin{subequations}
\label{eqn:EB}
\begin{eqnarray}
\left|\mathcal{E}\right| &=& \frac{M\left(r^3-3ra^2\cos^2\theta\right)}{\left(r^2+a^2\cos^2\theta\right)^{3}},\label{eqn:EB1} \\
\left|\mathcal{B}\right| &=& \frac{M\left(3r^2a\cos\theta-a^3\cos^3\theta\right)}{\left(r^2+a^2\cos^2\theta\right)^{3}},  \label{eqn:EB2}
\end{eqnarray}
\end{subequations}
and $\Theta_{\mathcal{E}} = \Theta_{\mathcal{B}} = 0$. This 
leads to the conclusion, using Eq.~(\ref{eqn:psi4tf}), that $\Psi_4 = 0$ in the
frame corresponding to $k=0$. 
Furthermore, we find the correct value for $\Psi_2=\frac{M}{\left(r+ia\cos\theta\right)^3}$,
which is exactly what we expect in the {\it Kinnersley frame}. We conclude
that the frame corresponding to $k=0$ converges to the {\it Kinnersley frame}
in the limit of a single Kerr hole; in other words, it is a {\it quasi-Kinnersley frame}. 

We emphasize
again that in our calculation only the expression for the time-like vector 
$N^a$ was given ``a priori''; no further assumptions on the other space-like vectors needed
to create the null tetrad were made. In fact, the correct identification of those space-like vectors
is implicitly achieved using the expressions for the Weyl scalars given in Eq.~(\ref{eqn:psi4tf})
and (\ref{eqn:psi2tf}).

\section{The weak field limit}
\label{sec:weakfield}

It is assumed here that the metric
is of the form $g_{ab}=\eta_{ab}+h_{ab}$ where $\eta_{ab}$ is the
flat Minkowski metric and $h_{ab}$ is the perturbation. 

In the limit of flat space-time, the $N^a$ vector of the {\it Kinnersley tetrad} 
given in Eq.~(\ref{eqn:normvector})
assumes the value $N^{a}=\left[\frac{3}{2\sqrt{2}},\frac{1}{2\sqrt{2}},0,0\right]$.
In relation to this vector, $E_{ab}$ and $B_{ab}$ are then given, using the transverse-traceless
gauge for the metric perturbation, by

\begin{subequations}
\label{eqn:EBweak}
\begin{eqnarray}
E_{\hat{\theta}\hat{\theta}} = -\frac{1}{2}\frac{\partial^2 h^{TT}_{\hat{\theta}\hat{\theta}}}{\partial t^2}, &
E_{\hat{\theta}\hat{\phi}} = -\frac{1}{2}\frac{\partial^2 h^{TT}_{\hat{\theta}\hat{\phi}}}{\partial t^2}, \\
B_{\hat{\theta}\hat{\theta}} = -\frac{1}{2}\frac{\partial^2 h^{TT}_{\hat{\theta}\hat{\phi}}}{\partial t^2}, &
B_{\hat{\theta}\hat{\phi}} = -\frac{1}{2}\frac{\partial^2 h^{TT}_{\hat{\theta}\hat{\theta}}}{\partial t^2}, 
\end{eqnarray}
\end{subequations}
where the hatted indices indicate contraction with the angular tetrad vectors.

The expressions for $\Psi_4$ and $\Psi_2$ given in Eq.~(\ref{eqn:psi4tf}) and
(\ref{eqn:psi2tf})
simplify considerably in a perturbed flat space-time. Specifically, 
the contribution to the curvature from the background
vanishes, which means that $\Psi_2$ must be zero. 
Using Eq.~(\ref{eqn:psi2tf}), one finds that in this case, $\Theta_{\mathcal{E}}=
\Theta_{\mathcal{B}}=\frac{\pi}{2}$
(assuming to have chosen $k=0$ corresponding to the transverse frame
which is also the {\it quasi-Kinnersley frame}, as shown in the previous section). 
Substituting this result into the expression
for $\Psi_4$ given in Eq.~(\ref{eqn:psi4tf}), one concludes that in this particular
limit,

\begin{equation}
\left(\Psi_4\right)_{QKF} = -\sqrt{\frac{E_{ab}E^{ab}}{2}}+
i\sqrt{\frac{B_{ab}B^{ab}}{2}}.
\label{eqn:psi4ertu} 
\end{equation}

It is interesting to notice that in the limit of perturbed flat space-time, the expression
found for $\Psi_4$ in the {\it quasi-Kinnersley} frame (Eq.~(\ref{eqn:psi4tf})) naturally converges to
a Poynting vector-like expression. More generally, the angles
$\Theta_{\mathcal{E}}$ and $\Theta_{\mathcal{B}}$ seem to play a fundamental role
in dividing the curvature expressed by the quantities $E_{ab}E^{ab}$
and $B_{ab}B^{ab}$ between background and gravitational waves. In this article, we have
in fact studied the two limits where these angles are either zero, so that the contribution to the curvature
is given by the background, or $\frac{\pi}{2}$, in which case only 
the gravitational radiation contributes to the curvature. In a general situation,
one expects $0<\Theta_{\mathcal{E}}<\frac{\pi}{2}$ and $0<\Theta_{\mathcal{B}}<\frac{\pi}{2}$, so that both the background and the gravitational
radiation contribute to the curvature. 

The expressions for $E_{ab}E^{ab}$ and $B_{ab}B^{ab}$ are given 
(neglecting quadratic and higher order terms) by 
$E_{ab}E^{ab} \approx \frac{1}{2}\left(\frac{\partial^2 h^{TT}_{\hat{\theta}\hat{\theta}}}{\partial t^2}\right)^2$
and 
$B_{ab}B^{ab} \approx \frac{1}{2}\left(\frac{\partial^2 h^{TT}_{\hat{\theta}\hat{\phi}}}{\partial t^2}\right)^2$,
leading to the final expression for $\Psi_4$:

\begin{equation}
\left(\Psi_4\right)_{QKF} \approx -\frac{1}{2} \left[\frac{\partial^2 h^{TT}_{\hat{\theta}\hat{\theta}}}{\partial t^2}-
i\frac{\partial^2 h^{TT}_{\hat{\theta}\hat{\phi}}}{\partial t^2}\right]. \label{eqn:psi4pertufin} 
\end{equation}

Eq.~(\ref{eqn:psi4pertufin}) shows that in the limit of a slightly perturbed flat space-time,
the expression of $\Psi_4$ is directly related to the two gravitational wave degrees of freedom
expressed in the $TT$ gauge. The results found in the previous two sections justify the use of Eq.~(\ref{eqn:psi4tf}) as a valid formula for wave extraction: it vanishes
for a single black hole space-time while it is related to the gravitational wave degrees of freedom
when dealing with a perturbed space-time.

%%%%%%%%%%%%%%%%%%%%%%
%%%   CONCLUSION   %%%
%%%%%%%%%%%%%%%%%%%%%%

\section{The $N^a$  time-like vector determination}
\label{sec:timelike}

In the previous sections, it has been shown that the choice of the correct tetrad, 
using the {\it quasi-Kinnersley frame} properties, can be reduced to the correct choice of
time-like vector used for the calculation of the electric and magnetic parts of the
Weyl tensor. Once this is done, the other vectors needed for the construction of the
null tetrad are easily identifiable 
and the expressions for the Weyl scalars found using
these vectors were given (Eq.~(\ref{eqn:psi4tf}) and (\ref{eqn:psi2tf})). 

The ``preferred'' time-like vector has been found to see the electric and magnetic
parts of the Weyl tensor with their eigenvectors aligned. Work is still in progress to determine
this vector in a general and robust way; however, a preliminary valid choice is that
of a time-like vector which converges to the right one asymptotically, which guarantees
that the expressions for the scalars are invariant at first order in perturbation theory. 

In a numerical simulation, the straightforward choice for the time-like vector is that
of the normal to the space-like Cauchy hypersurface. In the limit of  flat space-time,
such a vector converges to $N^a=[1,0,0,0]$ (assuming the lapse $\alpha\rightarrow1$ and the shift
$\beta^a\rightarrow0$
in this limit) which happens to be boosted in the flat limit with respect
to the $N^a$ found from the Kinnersley tetrad. The result of this
is that the factor $\frac{1}{2}$ in Eq.~(\ref{eqn:psi4pertufin}) 
disappears. This choice of the $N^a$ time-like
vector can be a good approximation. However, a way to correctly choose
$N^a$ in general must be found to rigorously complete this work.

%%%%%%%%%%%%%%%%%%%%%%
%%%   ACKNOWLEDGMENTS   %%%
%%%%%%%%%%%%%%%%%%%%%%

\acknowledgments

It is a pleasure to thank Luisa Buchman and Richard Matzner for careful proofreading of this manuscript.
I am also grateful to John Baker, James Bardeen and James van Meter for useful discussions.
This work
is funded by the NASA grant NNG04GL37G to the University of Texas
at Austin.

%%%%%%%%%%%%%%%%%%%%%%
%%%   REFERENCES   %%%
%%%%%%%%%%%%%%%%%%%%%%

%\bibliographystyle{apsrev}
\bibliography{references}

%%%%%%%%%%%%%%%
%%%   END   %%%
%%%%%%%%%%%%%%%

\end{document}